# Natural Typing Recognition via Surface Electromyography

Michael S. Crouch, Mingde Zheng, Michael S. Eggleston

*Abstract*—**By using a computer keyboard as a finger recording device, we construct the largest existing dataset for gesture recognition via surface electromyography (sEMG), and use deep learning to achieve over 90% character-level accuracy on reconstructing typed text entirely from measured muscle potentials. We prioritize the temporal structure of the EMG signal instead of the spatial structure of the electrode layout, using network architectures inspired by those used for real-time spoken language transcription. Our architecture recognizes the rapid movements of natural computer typing, which occur at irregular intervals and often overlap in time. The extensive size of our dataset also allows us to study gesture recognition after synthetically downgrading the spatial or temporal resolution, showing the system capabilities necessary for real-time gesture recognition.**

## I. Introduction

NATURAL gesture-based human-machine interfaces (HMI) require the ability to recognize gestures occurring at high speed and at irregular intervals. In the past decades, experimental systems using surface electromyography (sEMG) have been able to recognize a growing number of gestures, with increasing accuracy [1]–[4]; however, existing datasets and algorithms have still focused on recognizing one gesture at a time, typically within pre-defined time windows or at fixed time intervals [1], [5], [6]. These artificial limits are the biggest constraint to the deployment of sEMG-based natural interfaces.

Recent wearable gesture-recognition systems have used the techniques of *deep learning* to classify complex and fine-grained finger movements [4], [7], [8]. Early machine learning methods relied on human subject-matter experts to cleverly extract a few meaningful "features" from the signals to be classified, and then trained systems to recognize gestures based solely on these features [6], [9], [10]. In contrast, deep learning uses massive sets of training data to teach one neural network to perform end-to-end analysis, learning both which features of an input are important and how to classify inputs based on those features.

The quality of deep learning models, however, is limited by the size of the training data available; this is a fundamental obstacle in existing gesture recognition systems. Common practice is to record one gesture every 2–5 seconds in a highly controlled environment, not counting pauses between gestures or setup time. The largest published sEMG datasets contain fewer than 20,000 labeled recordings [8]. In contrast, modern deep learning techniques see continuing improvement with data sets of millions of images [11] or billions of words [12]. Even if such a monumental recording task were undertaken for gesture recognition, recording one gesture at a time does not accurately simulate natural human gestures, which are commonly overlapping in time and are performed at highly varying rates.

To overcome this challenge, we implemented a wearable system that captures forearm sEMG signals as a non-professional typist performs a natural typing task on a computer keyboard. Typing involves rapid, highly varying, fine-grained finger movement that can be performed by semi-skilled users. Average typing speed in a large sample of online survey volunteers was approximately 52 words per minute [13], corresponding to over 15,000 keystrokes per hour. Using the keyboard to automatically record and classify each character press, we generate a dataset consisting of over 148,000 labeled gestures, significantly larger than any sEMG gesture dataset previously reported. We then apply a novel deep learning-based AI model designed for real-time gesture recognition at naturally varying human rates, achieving almost 91% character-level accuracy for reconstructing typed text solely from forearm sEMG muscle activations.

The high accuracy of our real-time typing recognition model shows that it understands and predicts fine-grained finger motions at their natural speed. This model, and the dataset it is based on, will provide a starting point for transfer learning applications in HMI, prosthetics, and robotics. Our uniquely large training dataset also allows us to perform parametric requirement analyses, synthetically degrading data and retraining the model from scratch in order to probe the effect of lower spatial resolution or lower sampling rate on achievable accuracy. We obtain precise estimates for the hardware fidelity necessary for natural gesture recognition in both of these dimensions.

## II. Related Work

Modern computer interfaces, such as keyboards and touch

Michael S. Crouch, Mingde Zheng, and Michael S. Eggleston are with the Data & Devices Group, Nokia Bell Labs, 600 Mountain Ave, Murray Hill, NJ 07974. (email: {michael.crouch, mingde.zheng, michael.eggleston}@nokia-bell-labs.com})

screens, severely limit the large degrees of freedom and interaction complexity of natural human gestures. Recent work to capture more detailed human gestures is largely split between the use of video-based sensing techniques and wearable sensors [14]. Whereas video systems have shown impressive results, they are sensitive to background and lighting [15] and present privacy and surveillance concerns [16], which may preclude their general adoption.

The most widespread wearable sensors appropriate for human gesture recognition are inertial measurement units (IMUs), which measure acceleration and velocity. These are cost-effective and widely available, but work best for large arm motions, and are not well suited to capture fine motions of the fingers [17]. Surface Electromyography (sEMG), on the other hand, is a widely adopted wearable sensing technique that has shown increasing promise in sensing complex and fine-grained locomotion for gesture recognition. It has been explored for robotic control [18], prosthetics [19]–[22], and consumer-grade HMI [23], [24].

The human brain communicates with the body through an elaborate and extensive bioelectrical network. sEMG takes advantages of these pathways, using skin contact-based electrodes to measure the neuromuscular potentials used for voluntary control of the skeletal muscles. In gesture recognition applications, an armband is placed on the user's forearm, containing multiple electrodes which measure potential at different points (and thus track the activation of different muscles). Consumer grade sEMG devices are available but with limited capabilities; a notable example is the now-discontinued MyoWare armband [25].

Publicly available datasets of gestures and sEMG signals [3], [26]–[30] have fueled the continued development of deep learning techniques, but deployment of sEMG systems has still been constrained by the difficulty in inferring a large number of natural human gestures accurately and at natural gesture speed. This difficulty stems from the fact that control of each finger is a complex and coordinated activity of several different muscles [31], which makes simple classification of sEMG signals non-trivial. In addition, all implementations to date operate on very slow gesture frequencies on the order of one gesture per second or slower, significantly lower than most natural human gestures [32].

sEMG sensor input consists of a 1-dimensional or 2-dimensional spatial grid of electrodes sampling a signal that changes over time, forming a total of a 2-dimensional or 3-dimensional input. Early sEMG gesture recognition efforts discarded the temporal structure of the signals. Due to the physical mechanisms that generate sEMG signals, they are highly stochastic, and therefore it was historically accepted that the instantaneous value of an EMG signal was of little use [9], [29]. This approach separated each sEMG channel into fixed-length time windows, calculated a small number of fixed features over each window, and attempted to classify gestures based entirely on these features.

More recent work has used deep learning to focus on the spatial structure of sEMG signals, and has included an increasing number of electrode channels to help measure this structure. With many signal channels, a wide range of neural network techniques from image analysis can be employed. For instance, using sensor hardware with a 2-dimensional grid layout of electrodes ("high-density sEMG"), the spatial patterns in an instantaneous reading have been analyzed successfully with high accuracy via convolutional networks [1], [30]. Similar architectures have also been used for analyzing sparse sEMG signals through an entire time window, using a 2-dimensional convolution that looks for patterns over both the time and space axes [1], [4], [7], [33]–[35]. These spatial/temporal convolutions are sometimes combined with Long Short-Term Memory (LSTM) layers, for tracking longer-duration movements [34], [36]–[39]; however, these networks still separate input windows into fixed-size intervals and report classification at a constant rate, generally reporting one classification every 50-300 ms.

We believe that focusing on the spatial structure of sEMG signals, at the expense of understanding their temporal structure, has been a mistake. The large number of electrode channels, sometimes in excess of 100 [28], [30], [40], has led to expensive and complex custom-built systems. Different hardware systems have different spatial layouts; the use of convolutional neural networks in the spatial dimensions trains models which are specialized to the spatial structure of the wearable, limiting their ability to generalize or be used for transfer learning. But most importantly, we show that a per-channel architecture designed to prioritize the time structure of each channel leads to the ability to recognize gestures with high accuracy in real time.

The task of gesture recognition is much closer to transcribing spoken language than to analyzing video. Our methods are inspired by models for language transcription [41], with streaming networks that provide real-time analysis. These models take in a stream of data at a high frequency (either audio or sEMG) and produce a stream of classifications at a lower frequency (words or gestures). Training occurs on finite-length time windows, but in deployment the resulting network architecture can be evaluated indefinitely in real time, as readings occur.

## III. System Design

Our system is designed for ease of manufacture and data collection, and for power of generalization. Our neural network begins with a per-channel feature extraction block. By constructing the architecture to process each measurement channel separately and identically, the neural network is forced to learn a universal set of features for analyzing sEMG signals, which is important for enabling future reuse of the learned features with different physical hardware designs and parameters. The per-channel architecture was also chosen because sEMG signal noise is often concentrated among a small number of channels experiencing contact loss or motion artifacts [42]; by delaying the fusion of information from multiple channels, we ensure the network has access to as many "good" features as possible. By using the huge number of labeled typing events in our dataset, our network can model the time structure of sEMG signals, using convolutional layers for short-duration per-channel features and combining these with LSTM layers for tracking longer-duration multi-channel structure.

### A. *Wearable and Acquisition Hardware*

sEMG signals were captured using two custom-made compression sleeves tailored to fit the subject's upper forearms (Fig. 1; see Supplemental Material for complete fabrication information). Each sleeve contains 32 silver and silver chloride-based gel electrodes forming a total of 16 differential channel pairs per arm. The electrodes of each pair are separated by 5.1 cm along the proximodistal axis, parallel to the muscle fiber. Each pair is separated laterally by 1.3 cm, and successive pairs were offset in a "zigzag" pattern (see Fig. 1) by 2.5 cm.

sEMG signals were acquired using two synchronized iWorx RA-834 research stations, a dedicated commercially-available hardware for physiological data acquisition [43]. Equipped with four adjustable analog front-end amplifiers for biopotential signal conditioning, all 32 channels of sEMG inputs were sampled simultaneously in real time at a 16-bit resolution with a nominal system noise level less than 6 μV per channel. EMG signals have a typical power spectrum ranging from a few hertz up to 500 Hz and an unmagnified amplification ranging from a few hundred microvolts up to about 5 mV [44], [45]. The 4 analog amplifiers (the iWire Bio-8, also by iWorx) conditioned the acquired signal with an amplification factor of 1000, a 10 – 1000 Hz bandpass filter, and a sampling rate of 2000 Hz.

While sEMG signals were being recorded, a Python keystroke logger was run, recording keypresses by the typist and the corresponding timestamp of each. Code is available in Supplemental Materials. This logger ran on the same computer as the sEMG signal recording, for synchronization of timestamps. During the training data collection, the typist wore both sensor sleeves on the forearms, and typed on a standard USB keyboard in 20 – 30-minute sessions.

Because our measurement is of the forearm muscles, we are interested in characters which can be reached from the "home" typing position and therefore primarily rely on finger motion; however, we also wanted to include enough punctuation and special characters for natural transcription. We chose 32 characters: the letters 'a'-'z', space bar, period, comma, enter, backspace, and apostrophe. The keyboards used were a standard US (qwerty) layout.

Intersession variability is a major challenge in sEMG-based gesture recognition [46], [47]. Thus the sleeves were refit completely between sessions to acquire data from a variety of fittings. To reflect the natural typing task, the typist varied position and typing speed, and took occasional pauses. The resulting data averages 60 words per minute (wpm). Statistics of the collected dataset are presented in Table I and Fig. 2. The typist took dictation from episodes of the podcast "Jordan, Jesse, Go!", chosen for its natural conversation between native English speakers without a particular focus or topic, and attempted to type as much of the conversation as possible.

### B. *AI Model Architecture*

We present a neural network approach designed for real-time translation of data streams. The network is designed to take in a continuous stream of raw sEMG signals occurring at 2000 Hz and to output a continuous stream of the probabilities of corresponding keypresses at a natural typing rate. The deep learning network designed for this task can be understood as four main stages: (1) per-channel temporal down-sampling and feature extraction, (2) feature compression across multiple parallel channels, (3) long-term temporal feature extraction, and (4) keystroke classification.

Since our interest is in real-time input systems, we first focus on the temporal structure of the signals from each individual sEMG channel. Convolutional and pooling layers are used on each channel for feature extraction. The pooling layers make the network less sensitive to the exact timing of muscle activations, and also provide an effective down-sampling of the input frequency. This down-sampling can be seen as architecturally encoding our domain knowledge: that sEMG recording is happening at 2000 Hz, but that keystrokes will happen at a much lower frequency. Pooling and convolution lengths were chosen via experimentation, with the requirement of maintaining <100 ms processing time.

While each channel's features are calculated independently, the network extracts the same set of features from each channel. Different input channels cover different muscles in the arm and may thus see different signal patterns. By restricting the network to use the same features for each channel, we encourage it to learn a good general set of features for the analysis of sEMG signals, rather than learning features specialized for different channels. This increases the chance that the features learned will be suitable for reuse in different sEMG applications with different hardware configurations.

After the feature extraction and down-sampling section, the network has extracted 50 features for each of the 32 channels. The channel streams are merged to a flattened stream of 1600 features, occurring at 1/27th the frequency of the original 2000 Hz stream (~74 Hz). A 128-node Fully Connected (FC) layer is then used for data compression via dimensionality reduction; the same FC weights are applied at every timestep, yielding a stream of 128 features.

Following this dimensionality reduction, the third stage of our model is comprised of two Long Short-Term Memory (LSTM) layers. The features learned by the convolutional layers are limited by the receptive fields of the convolutions, which are constructed to be less than 100 ms to decrease input latency. Recurrent layers such as LSTMs are able to accumulate changes over longer periods of time, appropriately for modeling e.g. accumulated position changes due to muscle action. LSTM layers have shown to be better at learning long-term dependencies than other types of recurrent network [48]. The use of multiple LSTM layers was chosen because it assists the model in learning patterns at multiple timescales [41].

The LSTM layers yield a stream of 64 features occurring at about 74 Hz. Three FC layers are used to construct the final gesture classifier from these features. The final layer has 34 output nodes, representing the 32 character classes, plus a "character separator" class and a "no output" class used for the CTC metric. A SoftMax activation function is used after this layer, ensuring that the output can be interpreted as a probability distribution: the chance that each character has been recognized at each timestep.

Each convolutional layer and fully connected layer except the output layer is followed by a BatchNormalization layer and a ReLU activation. Dropout layers are used throughout for regularization. See the Supplemental Materials for the complete code and a complete textual description of all network parameters. The network is diagrammed in Fig. 3.

### C. *Offline Model Training*

Although the network is designed to run on continuous input data and to output inferences asynchronously, for training purposes it is necessary to separate the input recordings into finite-length segments. This segment length must be long enough that most keystrokes are well within the segment, so that errors from LSTM initialization are minimized, and so that the network's behavior during training approaches the behavior during real-time use. However, the segment length must also be short enough to fit within the GPU memory available.

For training purposes, the collected data was segmented into 15-second blocks; each block was associated with the series of keystrokes recorded during that block. The frequencies of the characters recorded are shown on Fig. 2. Partial blocks at the end of recording sessions were discarded. Any block containing a recorded keystroke not on Fig. 2 was discarded (27 blocks), leaving 1960 valid intervals. The *Connectionist Temporal Classification* (CTC) error function [49] was used with the Adam optimizer [50] with a learning rate of $10^{-3}$ to train the network for 300 epochs. Minibatches of size 32 were used across 4 GPUs.

The set of 1960 valid intervals was randomly shuffled, and the network was then trained using 10-fold cross validation. For each model trained, the prediction accuracy of the system is reported only on the holdout set. For each interval, CTC decoding with beam width five [49] was used to estimate the most probable string of keystrokes from the interval. This prediction was measured against the keyboard-recorded text using the Python "editdistance" package to determine the minimum number of insertions, deletions, and substitutions necessary to turn the predicted keystrokes into the recorded keystrokes.

## IV. Results & Discussion

During the 10-fold cross validation, the network achieved a mean character-level accuracy of 90.93%, a maximum accuracy of 91.63%, and a minimum accuracy of 90.11%. The system was thus able to achieve a very high accuracy for the recognition of the fine-grained finger gestures involved in typing, with a network architecture that allows application in real time.

The approximately 91% character-level accuracy of the system is already above the 87-89% pre-correction accuracy reported for mobile device touch-typing [51]–[53]. Examining the errors made by the system, below, shows that most errors involve two keys which are both adjacent on the keyboard and typed by the same fingers. This limited class of common errors should make spelling correction practical.

Our large training dataset also allows us to retrain the model after synthetically downgrading the temporal or spatial resolution, and study how these hardware characteristics affect the accuracy of the resulting system. This could help guide hardware requirements for future HMI systems. As expected, the accuracy achievable increases as the sampling frequency increases and as the number of channels available increases. In particular, spatial down-sampling shows that approximately 90% accuracy is already achieved at 12 channels per arm, suggesting that our focus on the temporal structure of the signals allows significantly simpler hardware construction.

### A. *Error Modeling*

During training, our network predicts a string of characters for each training interval; however, because typed characters do not happen at predictable intervals, there is not a one-to-one correspondence between characters typed and characters predicted. This is critical to the model's ability to interpret a varying speed of typing in real-time; however, it also makes it more complicated to understand exactly what errors the model is making.

The mathematical problem of examining a set of inputs and outputs and determining the implicit probabilities of various types of errors has been studied in the field of spelling correction. By modeling the system as a noisy channel capable of making certain classes of errors [54]–[56], we can calculate what error probabilities would make the observed strings most likely (the "maximum likelihood" probabilities). We adapt the method of [57], analyzing a channel containing erroneous deletions, insertions, and substitutions [58] and applying the Expectation Maximization (EM) algorithm [59]. Possible errors are modeled as follows:

1. Deletion errors: each character $x$ has a different chance $p_{\text{del}}(x)$ of being deleted.
2. Insertion errors: there is a chance $P_{\text{ins}}$ that before transmitting a character, the channel first inserts an erroneous character. This may happen repeatedly. When a character is inserted, it has a different chance $p_{\text{ins}}(x)$ to be each character $x$. Note $\sum_x p_{\text{ins}}(x) = 1$.
3. Substitution errors: When a character $x$ is transmitted through the channel, for each possible character $y$, there is a chance $p_{\text{sub}}(x, y)$ that $y$ will be received.

Note that for each character $x$ we have

$$p_{\text{del}}(x) + \sum_y p_{\text{sub}}(x, y) = 1\,. \qquad (1)$$

To calculate confusion matrices, we apply this algorithm on the total set of predictions (training set and test set) from all ten training runs of the network. The resulting probabilities are shown in Fig. 4, Fig. 5, and Fig 6.

### B. *Errors Made*

We examine keystrokes of the left hand in detail; the right hand is similar, except as noted. The index finger is responsible for the largest number of keys: 'r', 't', 'f', 'g', 'v', and 'b'. These keys are infrequently missed, with $p_{\text{del}} < 0.01$ except for the 'r' key which has $p_{\text{del}}(\text{r}) = 0.0157$. Nearly all substitution mistakes are for other keys using the same finger, and typically between pairs of adjacent keys, including {r, t}, {f, g}, {f, v}, and {g, b}. The least likely key to be recognized as itself is 'v', with $p_{sub}$(v, v) = 0.797. This may reflect the relative rarity of this key within this natural language dataset, f(v) = 0.0080, but note that 'v' is almost always recognized as some other keystroke, with $p_{del}$(v) = 0.0084. We suspect that the finger motion is detected well, but that the measured forearm muscle activation does not capture whole-hand motion out of the home position, which primarily occurs in the shoulder and biceps.

The middle finger is responsible for the keys 'e', 'd', and 'c'. Again, we observe that all three keystrokes are more than 99% likely to be detected, with $p_{del}$ < 0.01, and that most substitution mistakes are for adjacent pairs. We also see a strong asymmetry between characters that occur with different frequencies (note f(e) = 0.0898, f(d) = 0.0269, and f(c) = 0.0176). For example, the character 'd' is 6.4% likely to be recognized as the more common 'e', whereas 'e' is only 1.8% likely to be recognized as 'd'. Similarly, 'c' is recognized as the more common 'd' 8.8% of the time, whereas 'd' is recognized as 'c' only 3.1% of the time.

The left ring finger covers the 'w', 's', and 'x' keys. There is a wide range of frequencies, with 's' very common and 'x' very uncommon, with f(s) = 0.0484, f(w) = 0.0186, and f(x)=0.0014. The 'x' key is recognized as a keystroke only 97% of the time, $p_{del}$(x)=0.0299, and is substituted with the common 's' fully 22.5% of the time (the highest substitution rate of any key pair).

The left pinkie covers the 'q', 'a', and 'z' keys. In this case, 'a' is very common and the other two are very uncommon, with f(a) = 0.0606, f(z) = 0.0011, and f(q) = 0.0010. These two keys are the most likely keystrokes to be discarded, with $p_{del}$(q) = 0.0761 and $p_{del}$(z) = 0.0672, and each have almost 10% chance of being misrecognized as 'a'. 'z' also has a 7.3% chance of being misrecognized as 's', showing some confusion between fingers.

Results for the right hand are similar, and we will highlight only the differences. For the right index finger, we see larger chances of substitution errors than for the left index finger, perhaps because the bottom row index finger keys are farther from the home position for the left hand ('v' and 'b') than for the right hand ('n' and 'm'), and thus are more distinct. There is slightly more confusion between different fingers of the right hand than for the left hand.

The right-hand pinkie is responsible for several characters: 'p', the apostrophe, the backspace key, and the enter key. The 'p' and apostrophe keys show similarly poor recognition to the left-hand pinkie keys 'q' and 'z', even though they are much more common, with f(p) = 0.0125 and f(') = 0.0084. The backspace and enter keys show good recognition accuracy, however, possibly because they involve much more movement.

The space bar was typed by the typist entirely using the right thumb. Accuracy for the space bar was very high, as expected because it is the only character typed with that finger, and because it is by far the most common character in the dataset (over 18% of typed characters).

Overall, we see that most substitution errors occur between two characters which are both adjacent on the keyboard and typed by the same finger. A confusion matrix for classifying only which fingers were used is shown in Fig. 7 and Fig. 8; it shows greater than 94% accuracy for each finger. Combined with the relatively low chances of insertion and deletion errors, and the predictable structure of these as well, we believe that distribution-aware spelling correction would make the system eminently usable in practice.

We observed above several instances where rarer characters have higher deletion chances, and where substitutions between adjacent characters are biased towards the more frequent character in the pair. Poor recognition results for rare characters may arise from a combination of factors. First, there are a limited number of occurrences of these characters in the dataset. Secondly, our system is trained on natural English text, and is expected to learn quickly to encode the marginal probabilities that each character will occur. The system may simply demand "stronger evidence" of 'z' than of 'a', because 'z' is a priori less likely to be typed. It is unclear which of these effects dominates the error, and thus how much recognition could be improved by additional data collection.

### *C. Spatial Resolution Requirements*

To estimate the number of electrode channels necessary for gesture recognition, the network was retrained with spatial resolution synthetically downgraded via linear interpolation. Data was originally recorded with 16 channels of recording electrodes per arm (numbered 0 through 15). To down-sample this data to $k$ channels, we chose channel 0, channel 15, and ($k$-2) equally spaced channels between them. For example, for $k$=5, the channel positions {0, 3.75, 7.5, 11.25, 15} were chosen; 3.75 is a synthetically constructed channel comprising 0.25 times the value of channel 3 plus 0.75 times the value of channel 4. For the special case of $k$=1, the only channel used is 7.5, the average of the middle two channels on the brace.

Our per-channel architecture allowed us to perform this degradation without making any other changes to the network or hyperparameters. Besides the altered number of input channels, the network, training process, and hyperparameters used were completely identical to the original $k$=16 case; the network was retrained from scratch, using only the down-sampled data. Results are shown in Fig. 9; numerical results available in Supplemental Materials

We see the expected increase in accuracy as the number of channels increases; however, the number of channels necessary for accuracy is much lower than reported in previous work using smaller datasets and spatially-structured neural networks [28], [60]. The system reaches an average 84.8% accuracy with only $k$=4 channels, and 88.2% accuracy with $k$=6 channels. Note that for $k$ in the set {2, 4, 6, 16}, no synthetic channels are used, ruling out the possibility that our linear interpolation methods have somehow augmented the information available to the network. We thus conclude that real-time gesture recognition is achievable using significantly fewer data acquisition channels than typically used in current experimental systems, when a sufficiently large set of training data is available.

### D. Temporal Resolution Requirements

The per-channel architecture of the system made it easy to synthetically downgrade the spatial resolution, by discarding channels and using linear interpolation. Studying the temporal resolution necessary for gesture recognition is more complicated. The network, as built, includes three pooling layers, each of which down-samples the input stream by a factor of 3. The resulting 27x reduction in frequency was chosen partially based on domain knowledge: that the input occurs at 2000 Hz, and that an output of approximately 74 Hz should be frequent enough to report register nearby pairs of keystrokes. If we were to simply discard half of the points, the temporal range of each convolution would double; the expressivity of the model would not be the same.

To preserve the network architecture, we instead synthetically down-sample each recording session to a lower frequency using the function `signal.resample()` from the Python `scipy` package. We then up-sample back to 2000 Hz so that the network architecture can be retained. At each frequency, the network is then retrained from scratch, using exactly the same architecture, hyperparameters, and training code. The resultant gesture recognition accuracy is shown in Fig. 10 (numerical results in Supplemental Materials).

Our results on the necessary temporal resolution largely agree with previous literature. Because the power spectrum of sEMG signals is most dominant up to the 500 Hz frequency band, a Nyquist sampling frequency of 1000 Hz is sufficient for avoiding aliasing effects [45]. We report an 85% accuracy at a surprisingly low sampling rate (200 Hz) using our model system, but then show incremental improvement in accuracy until at least 1000 Hz. This may explain the range of values reported in the literature for necessary sampling rates, which has included 200 Hz [61], 400 Hz [62], and 600 Hz [60].

## V. Conclusion & Future Outlook

We report the development of a system that recognizes fine-grained finger gestures in real time, using only the electrical signals measured by 16 sEMG channels on each forearm. Our system achieves over 90% character-level accuracy, without any data preprocessing, postprocessing, or spelling correction.

By focusing on the temporal structure of each sEMG channel, rather than the spatial structure of the measurement system, we design a neural network that is capable of low-latency recognition of irregularly-timed gestures occurring much more quickly than the 1–3 second gestures most commonly studied in the gesture recognition community. The use of a keyboard as a natural gesture recording device allows us to collect the largest sEMG gesture recognition dataset available, which we hope will be of further independent use.

Examination of keystroke recognition errors indicates that most involve difficulty identifying which of two adjacent keys a particular finger is pressing. For achieving high pre-correction accuracy, we believe it is necessary to integrate an Inertial Measurement Unit to measure hand or arm acceleration and positioning. Nonetheless, the extremely regular nature of the errors helps us conclude that spelling correction should be eminently possible.

Further improvement to our gesture recognition system would require intersession analysis, interuser testing, and validation beyond what this work has yet achieved. However, we have shown that with enough training data, our networks can learn the time structure of sEMG signals, and that achieving high-accuracy real-time gesture recognition is possible.

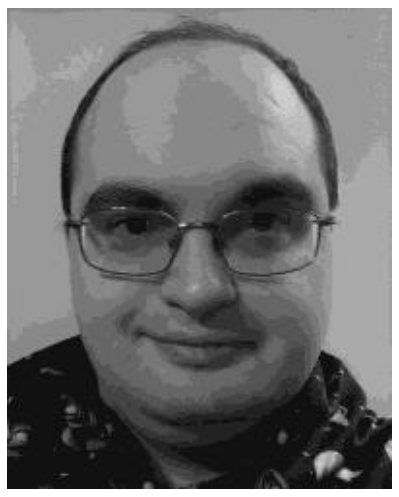

**Michael S. Crouch** received B.S. degrees in Physics and Theories of Computation from Carnegie Mellon University in Pittsburgh, PA, USA and his M.S. and Ph.D. in Computer Science from the University of Massachusetts, Amherst, USA, with a focus on algorithms for large data streams. He became a research scientist at Nokia Bell Labs in Dublin, Ireland in 2013, and has been at Murray Hill, NJ since 2016. His research focus is low-resource analytics and deep learning for high-frequency time series data; his interests include signal processing, learning for wearables, and biological data streams.

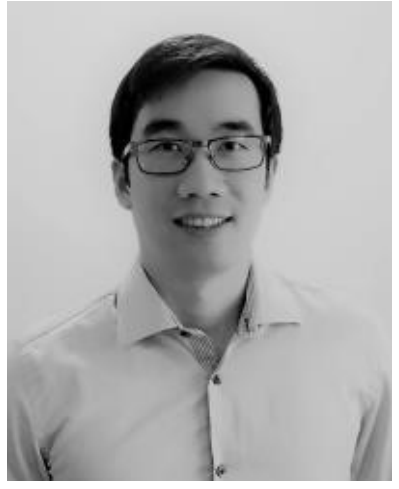

**Mingde Zheng** received his B.S. degree in Biomedical Engineering from the University of Texas at Austin, TX, USA in 2010. Later, he received his Master of Engineering and Ph.D. degrees in Biomedical Engineering with focuses on developing cellular sensors and Lab-on-Chip devices from Rutgers, the State University of New Jersey at New Brunswick, USA. Since 2016, Mingde has been a senior research scientist at Nokia Bell Labs, Murray Hill, NJ overseeing the development of non-invasive bioelectrical technologies for diagnostic and human-machine interaction applications. His research interests include inventing novel biosensors and actuators, point-of-care diagnostics, and functional biomaterials for human-center investigations. Mingde was the recipient of multiple NSF graduate fellowships and academic excellence awards, and he holds five patents on novel bioelectric sensing and stimulation technologies.

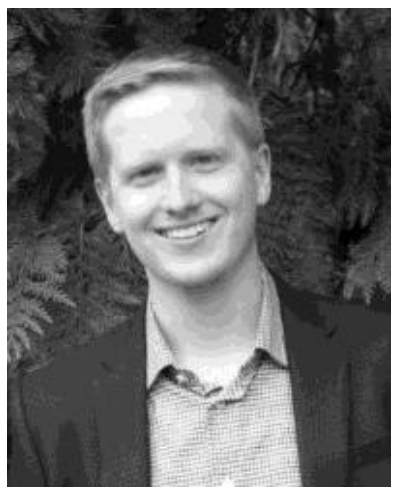

**Michael S. Eggleston** received the B.S. degree in electrical engineering and physics from Iowa State University, IA, USA and the Ph.D. degree in electrical engineering from University of California, Berkeley, CA, USA. In 2015, he joined Nokia Bell Labs in Murray Hill, NJ where he currently leads the Data & Devices Research Group. This interdisciplinary team is building the devices and AI systems that will seamlessly connect the physical and digital worlds of the future. His research interests include non-invasive human sensing, energy-autonomous devices, and AI for multi-modal perception. Michael holds 10 patents on integrated photonic and sensing technologies.

TABLE I
DATASET STATISTICS

| | |
|---|---|
| Keystrokes | 148,739 |
| Characters | 32 |
| Duration | 29,565 s<br>8 h 12 m 45 s |
| Channels | 32 |
| Sample Frequency | 2000 Hz |
| Data Points | 1,892,160,000 |
| Keystroke / s | ~5.03 |
| Average WPM | ~60.37 |

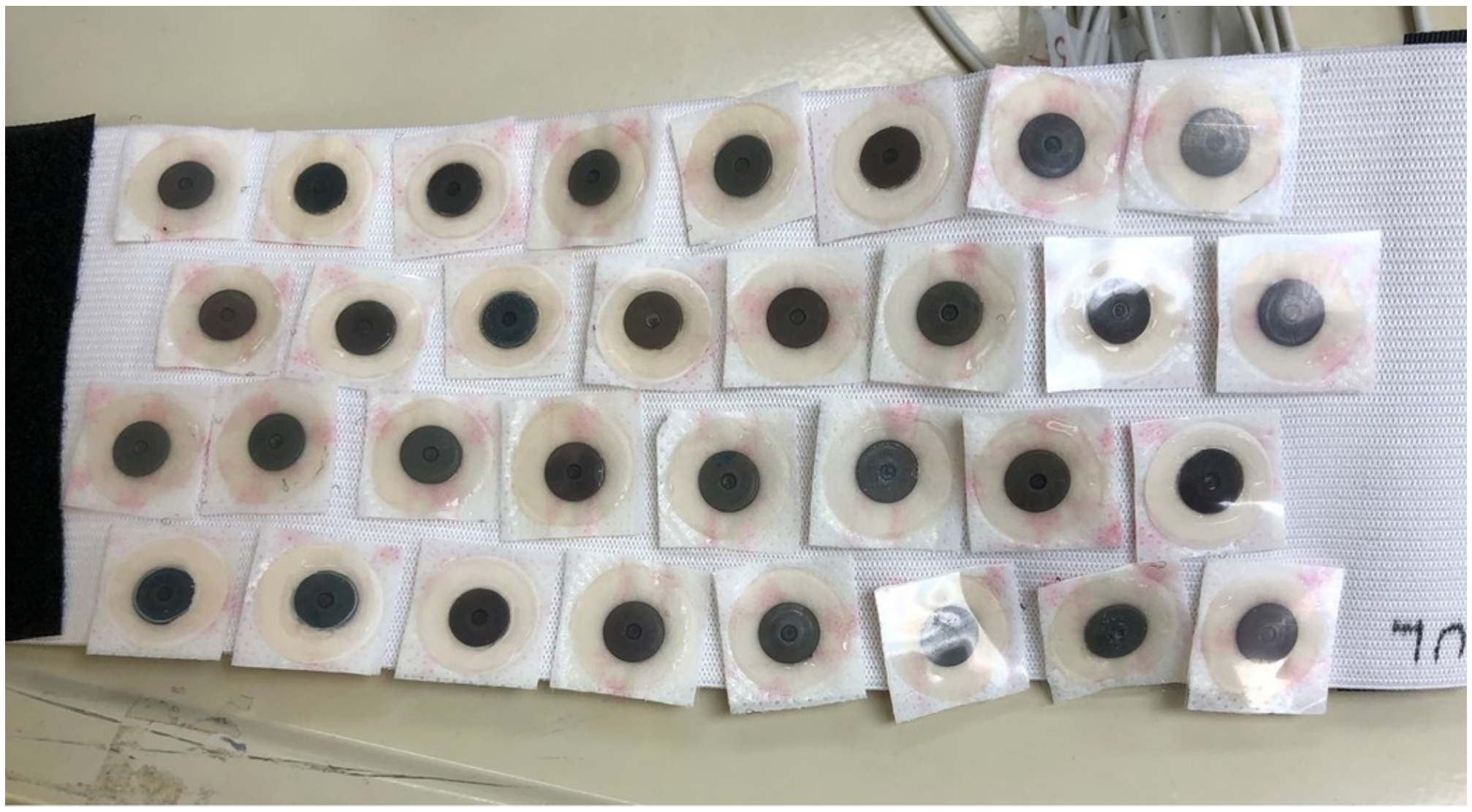

Fig. 1. Constructed arm brace.

88x48mm (300 x 300 DPI)

TABLE I
DATASET STATISTICS

| | |
|---|---|
| Keystrokes | 148,739 |
| Characters | 32 |
| Duration | 29,565 s<br>8 h 12 m 45 s |
| Channels | 32 |
| Sample Frequency | 2000 Hz |
| Data Points | 1,892,160,000 |
| Keystroke / s | ~5.03 |
| Average WPM | ~60.37 |

Table I.
Dataset Statistics.

225x105mm (281 x 281 DPI)

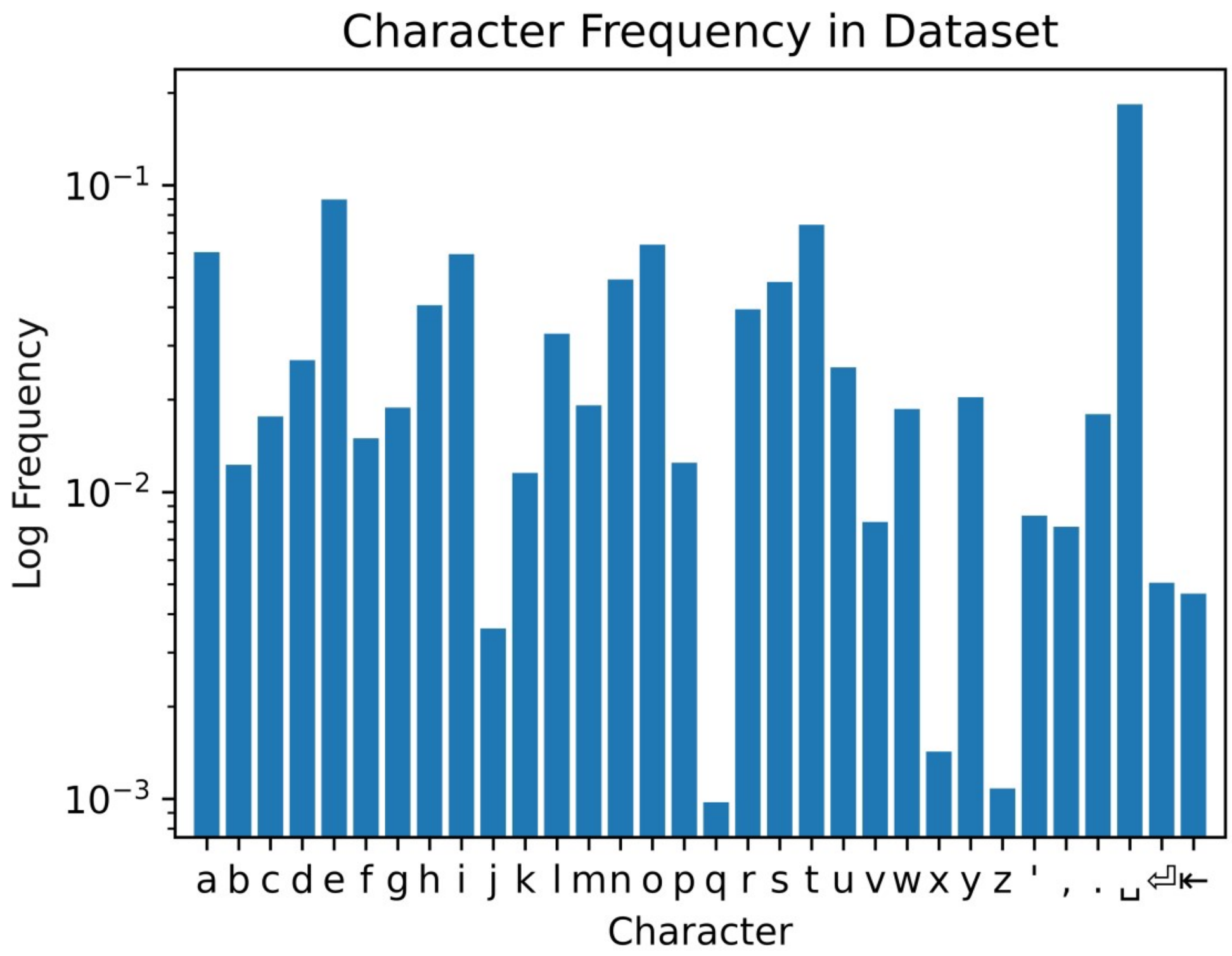

Fig. 2. Log frequency for each character in dataset. Numerical counts in Supplemental Material.

302x236mm (300 x 300 DPI)

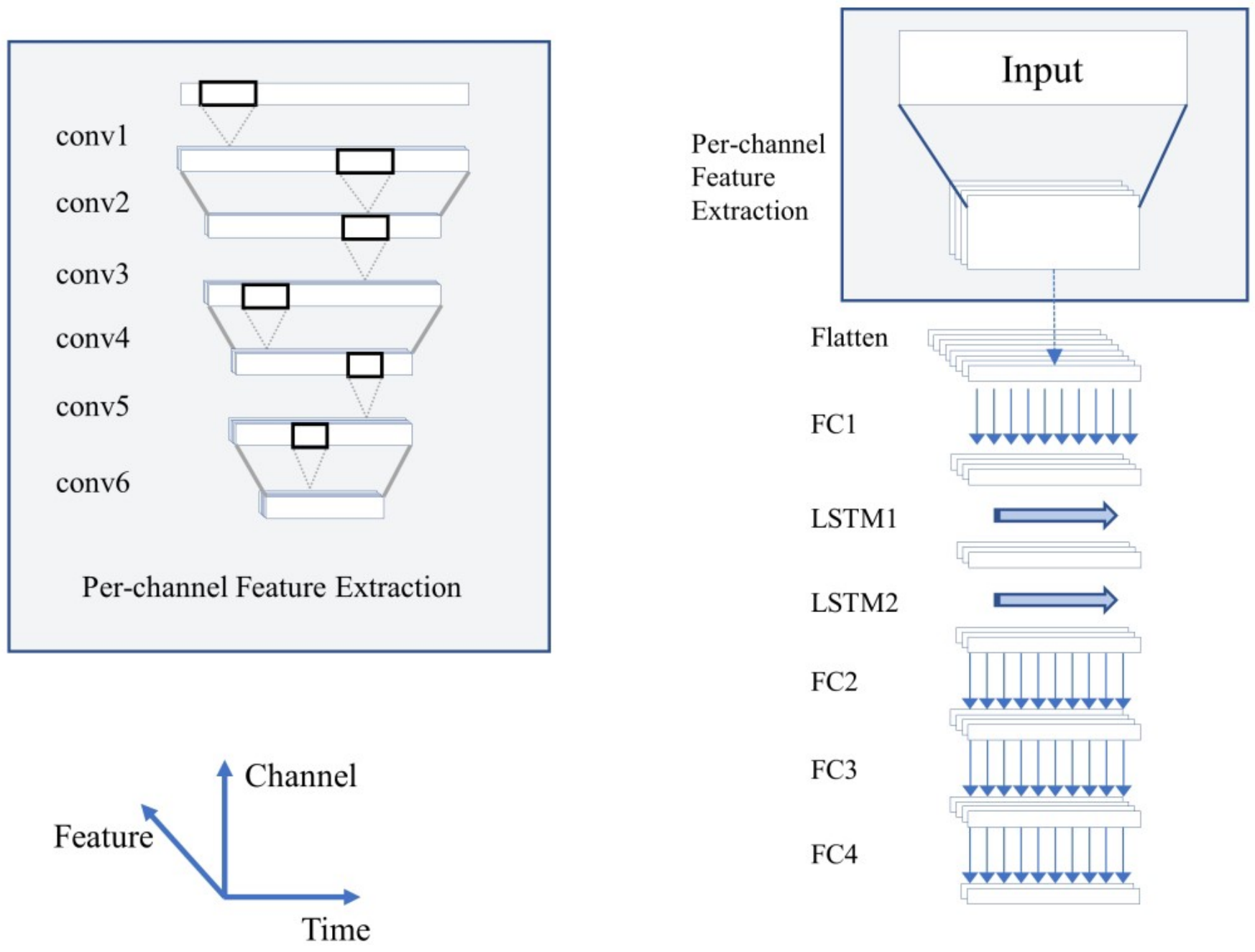

Fig. 3. Diagram of LSTM-CNN network, acting on one block of 30,000 samples. Not pictured: Dropout layers, Activation layers, BatchNormalization layers (see text).

463x323mm (157 x 157 DPI)

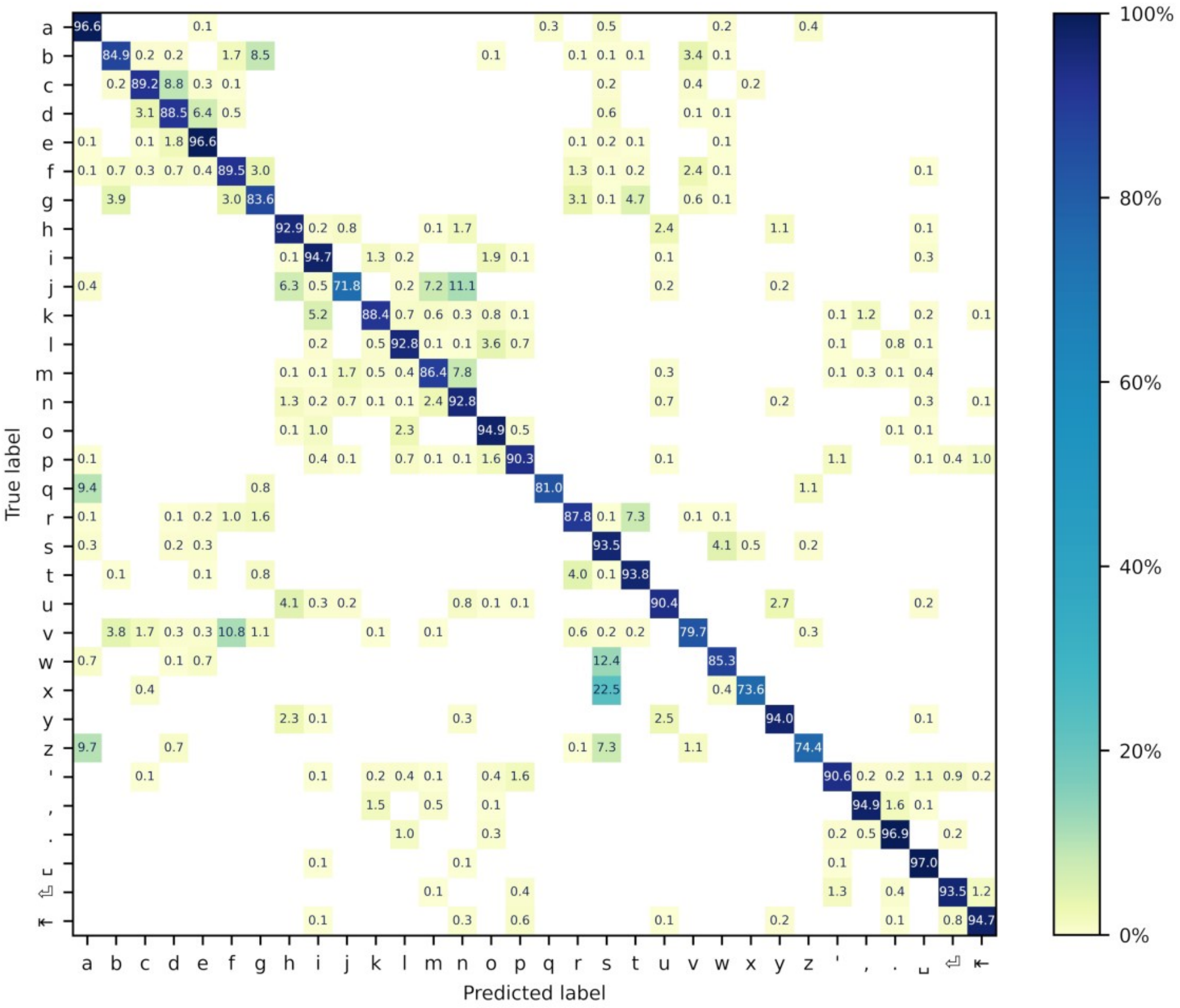

Confusion matrix (%) for substitution errors. When predicted and true labels are equal, this models the percentage of characters which are typed on the keyboard and which appear correctly in the neural network output. When predicted and true labels differ, this models the percentage of characters which are recorded by the keyboard as the "true label", are recognized as keystrokes by the neural network, but are then misclassified as the "predicted label". Note that rows do not add to 1 because of the possibility of character deletion (Fig. 7).

230x198mm (300 x 300 DPI)

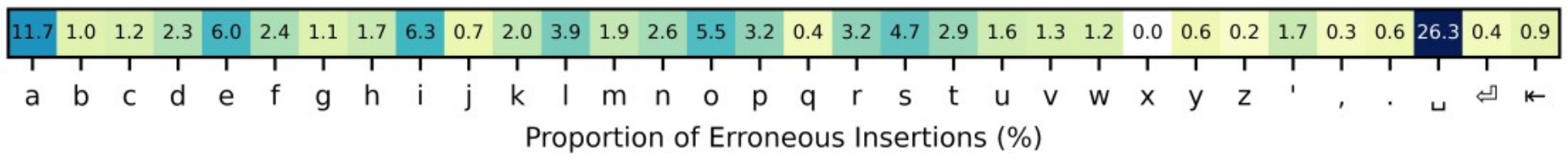

Fig. 5. Chance of erroneously inserting a character into the predicted text that was not in the ground text. This models characters which were predicted by the neural network but were not recorded by the keyboard. Chance is given as a proportion of erroneous insertions (~1.36% chance).

461x57mm (300 x 300 DPI)

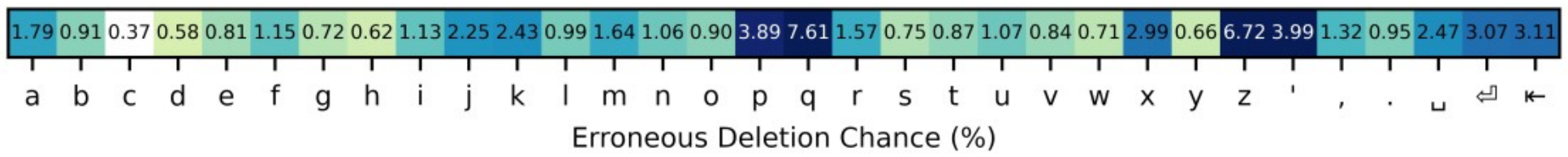

Fig. 6. Chance that a character will be erroneously deleted from the ground text. This models the chance that a character will be recorded by the keyboard but not predicted by the neural network.

461x57mm (300 x 300 DPI)

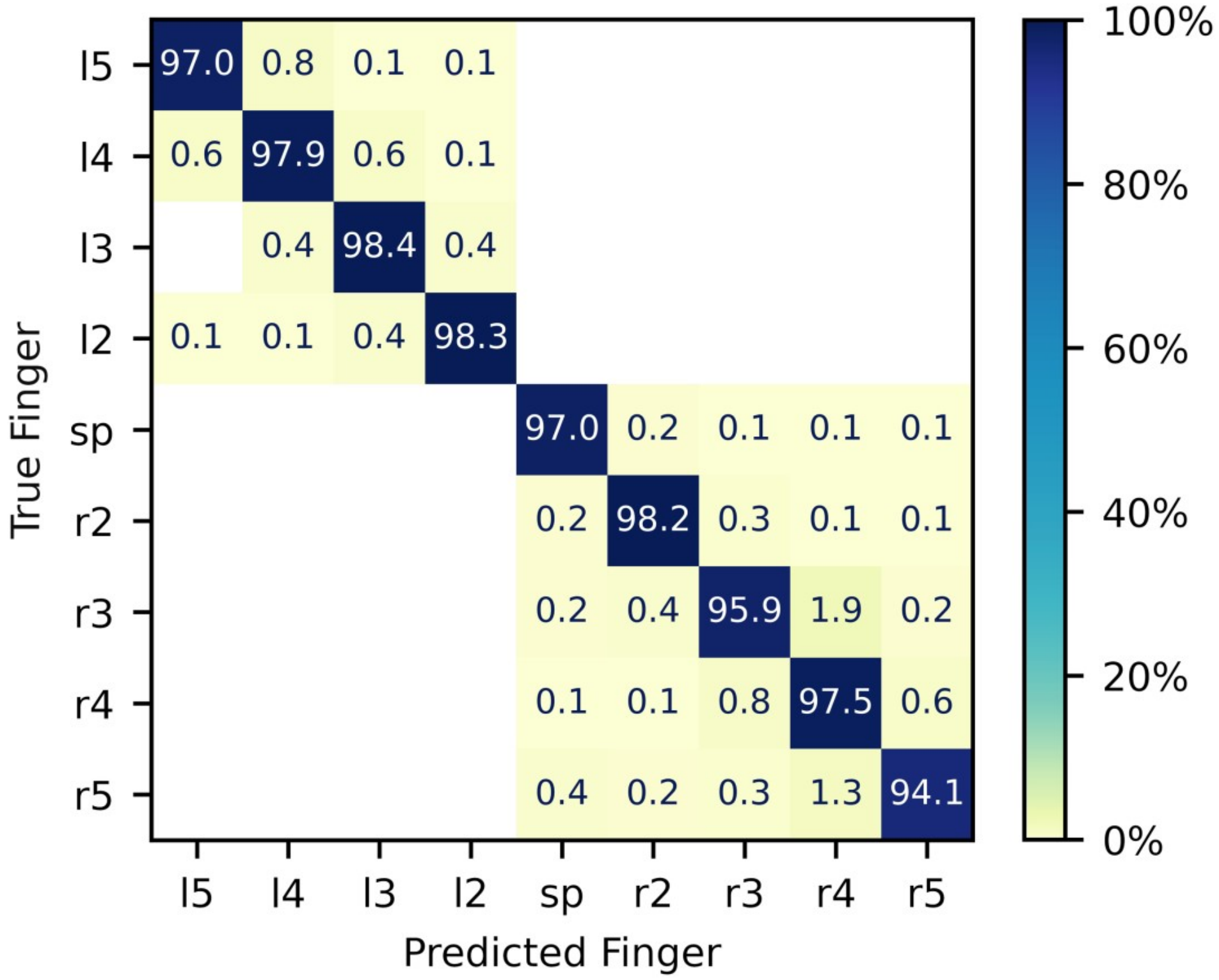

Fig. 7. Per-finger confusion matrix. Models the chance that a keystroke by True Finger is typed on the keyboard but that the neural network predicts some key corresponding to Predicted Finger. Fingers are numbered from 2 (index finger) to 5 (pinkie), for left and right hand. Right hand thumb denoted "sp" for space bar. Left hand thumb not used.

225x184mm (300 x 300 DPI)

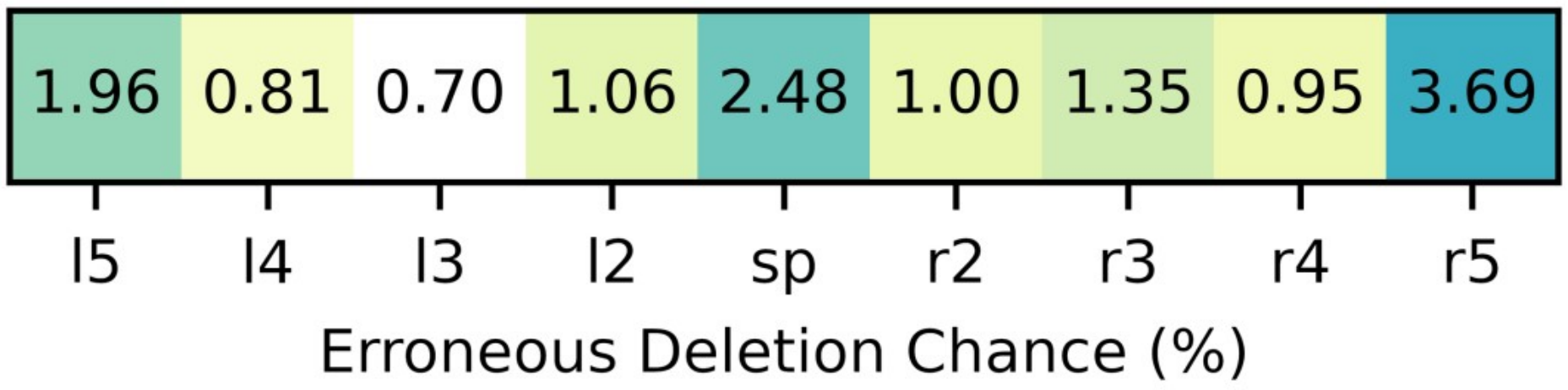

Fig. 8. Per-finger deletion matrix. Models the chance that a keystroke is recorded by the keyboard but not recognized by the neural network, conditioned on which finger is used for the key. Finger numbering as per Fig. 7.

225x65mm (300 x 300 DPI)

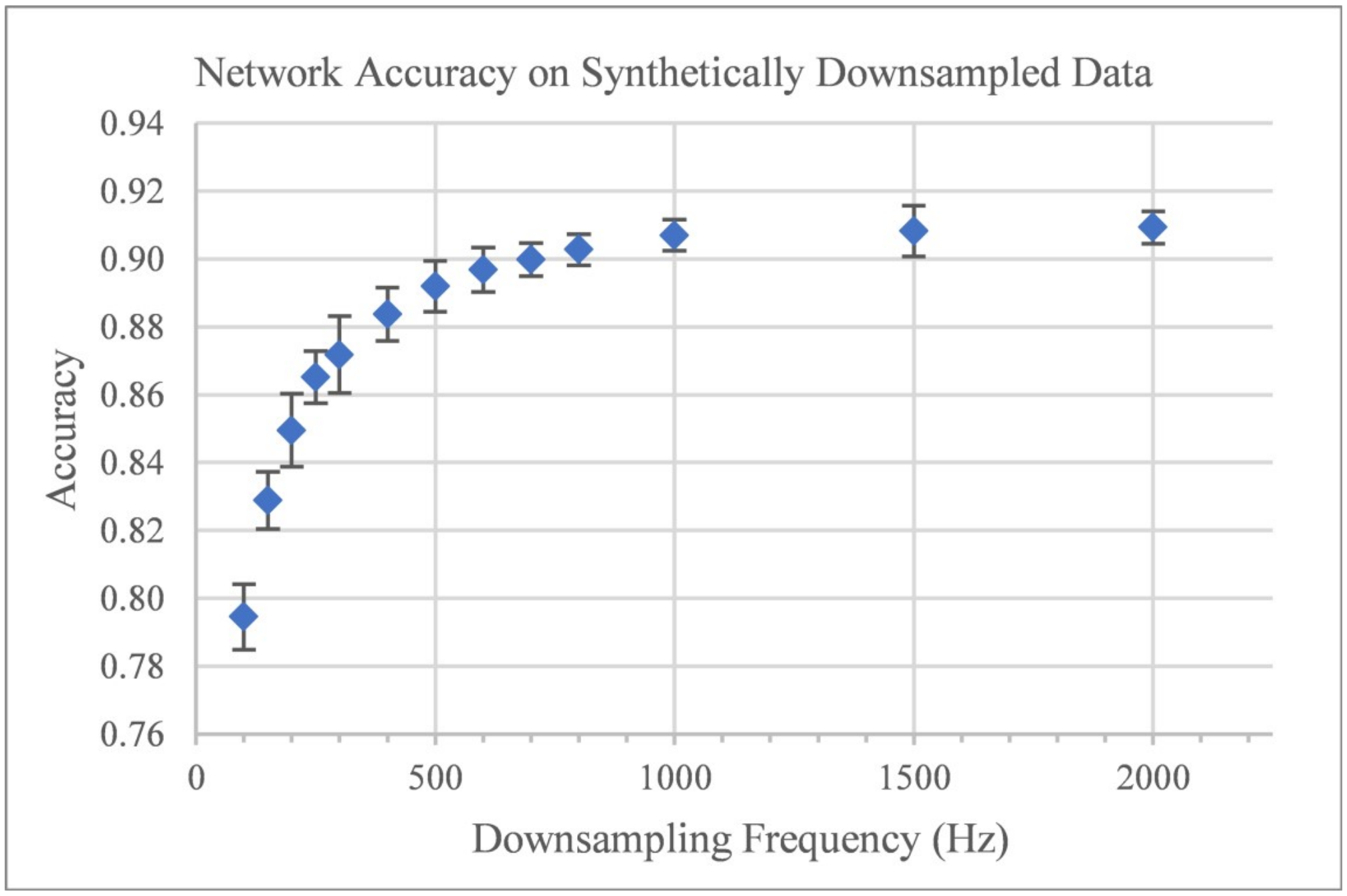

Fig. 9. Mean accuracy of synthetically down-sampled dataset over 10-fold cross-validation. Accuracy calculated on holdout sets. Error bars show standard deviation over the 10 runs.

225x149mm (300 x 300 DPI)

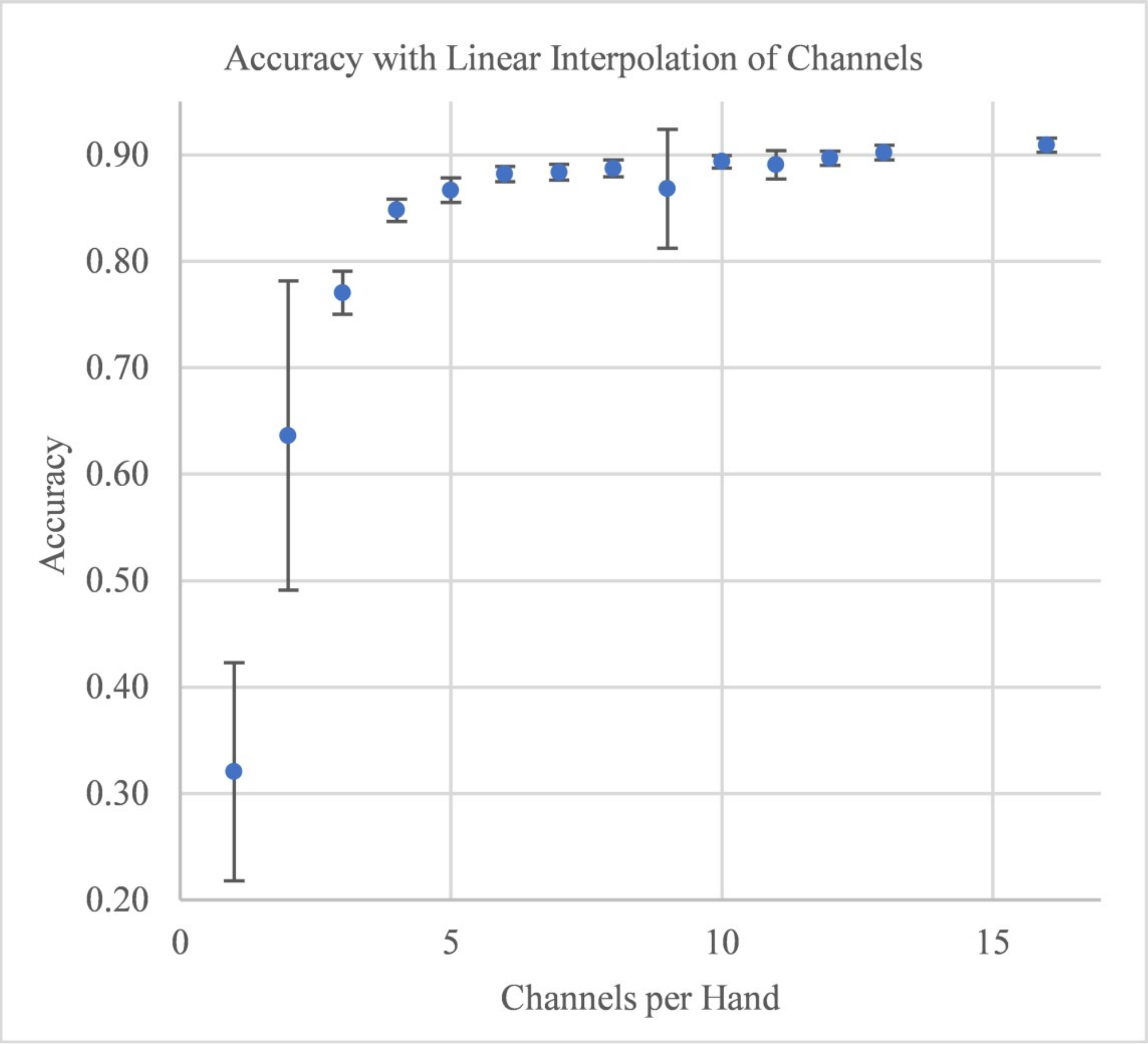

Fig. 10. Mean accuracy of spatially resampled dataset over 10-fold cross-validation. Accuracy calculated on holdout sets. Error bars show standard deviation over the 10 runs. Note that one run for k=9 did not converge in the 300 training epochs available; the other 9 runs followed the expected pattern.

225x205mm (300 x 300 DPI)